\documentclass[reprint,prl,amsmath,twocolumn,nofootinbib]{revtex4-1}
\usepackage{graphicx,epsfig}
\usepackage{xcolor}

\begin{document}

\author{Ming-Guang Hu}
\author{ Michael J. Van de Graaff}
\author{Dhruv Kedar}
\author{John P. Corson}
\author{Eric A. Cornell}
\author{Deborah S. Jin}

\affiliation{JILA, NIST and University of Colorado, Boulder, CO 80309, USA}
\affiliation{Department of Physics, University of
Colorado, Boulder, CO 80309, USA}

\title{Bose polarons in the strongly interacting regime }
\date{\today}

\begin{abstract}
When an impurity is immersed in a Bose-Einstein condensate, impurity-boson interactions are expected to dress the impurity into a quasiparticle, the Bose polaron.  We superimpose an ultracold atomic gas of $^{87}$Rb with a much lower density gas of fermionic $^{40}$K impurities.  Through the use of a Feshbach resonance and RF spectroscopy, we characterize the energy, spectral width and lifetime of the resultant polaron on both the attractive and the repulsive branches in the strongly interacting regime. The width of the polaron in the attractive branch is narrow compared to its binding energy, even as the two-body scattering length formally diverges.
\end{abstract}

\maketitle

\begin{figure}
\includegraphics[width=3.4 in]{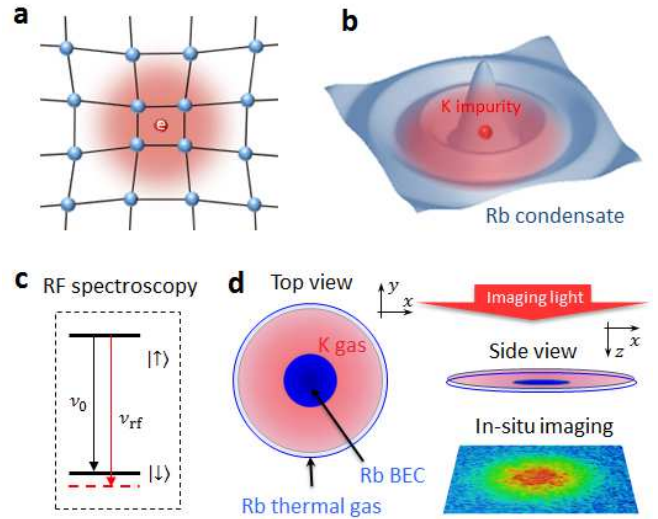}
\caption{Impurities immersed in a bosonic bath. \textbf{a}, Illustrations of the Bose polaron formed by an electron moving in a crystal lattice and \textbf{b}, its counterpart of an impurity in a BEC. \textbf{c},  RF spectroscopy of $^{40}$K impurities in a $^{87}$Rb BEC. The black lines denote two hyperfine states of bare K atoms and the red dashed line is the shifted energy level due to interactions with the BEC. \textbf{d}, Geometry of the trapped BEC and impurity clouds. The dark blue represents the Rb BEC cloud, the light blue shows the Rb thermal cloud, and the red shows the K impurity cloud. The imaging light propagates from top to bottom along $z$.}
\label{fig:Bosepolaron}
\end{figure}

The behavior of a dilute impurity interacting with quantum bath is a simplified yet nontrivial many-body model system with wide relevance to material systems.  For example, an electron moving in an ionic crystal lattice is dressed by coupling to phonons and forms a quasiparticle known as a Bose polaron (see Fig. \ref{fig:Bosepolaron}\textbf{a}) that is an important paradigm in quantum many-body physics \cite{Mahan}.   Impurity atoms immersed in a degenerate bosonic or fermionic  atomic gas are a convenient experimental realization for  Bose or Fermi polaron physics, respectively. 
Recent theoretical work \cite{Astrakharchik2004,Cucchietti2006,Kalas2006,Bruderer2008,Huang2009,Tempere2009,Volosniev2015, Grusdt2015} has explored the Bose polaron case, and the ability to use a Feshbach resonance to tune \cite{Chin2010}  the impurity-boson scattering length $a_\text{IB}$ opens the possibility of exploring the Bose polaron in the strongly interacting regime \cite{Rath2013,Li2014,Pena2015,Levinsen2015}.   Experiments to date \cite{Palzer2009,Catani2012,Spethmann2012,Scelle2013,Fukuhara2013,Kurn2014}  have focused on the weak Bose polaron limit.  The Bose polaron in the strongly interacting regime is interesting in part because it represents step towards understanding a fully strongly interacting Bose system.  While $a_\text{IB}$ can be tuned to approach infinity, the boson-boson scattering length $a_\text{BB}$ can still correspond to the mean-field limit.    A dilute impurity interacting very strongly with a Bose gas that is otherwise in the mean-field regime is, on the one hand, something more difficult to model and to measure than a weakly interacting system. On the other hand it is theoretically more tractable, and empirically  more stable than a single-component ``unitary" Bose gas in which $a_\text{BB}$ diverges and thus every pair of atoms is strongly coupled \cite{Makotyn2014}.

Our experiment employs techniques similar to those used in recent Fermi polaron measurements \cite{Schirotzek2009,Kohstall2012,Koschorreck2012,Zhang2012}. However, there are important differences between the Bose polaron and the Fermi polaron.  From a theory point of view, the Bose polaron problem involves an interacting superfluid environment and also has the possibility of three-body interactions \cite{Levinsen2015}, both of which are not present for the Fermi polaron. And on the experimental side, both three-body inelastic collisions and the relatively small spatial extent of a BEC (compared to that of the impurity gas) create challenges for measurements of the Bose polaron.
This work, in parallel with work done at Aarhus \cite{Jorgensen2016}, describes the first experiments performed on Bose polarons in the strongly interacting regime. In our case, the impurity is fermionic, not bosonic.   We report measurements of the Bose polaron energies and lifetimes using RF spectroscopy 
of fermionic $^{40}$K impurities in a BEC of $^{87}$Rb atoms. We tune the impurity-boson interactions using a Feshbach resonance, and our measurements reveal both an attractive and a repulsive polaron branch, whose energies agree with recent predictions \cite{Rath2013,Li2014,Pena2015}. We find that the Bose polaron exists across the strongly interacting regime and has a larger binding energy than does the low-density, two-body molecular state. 

\begin{figure*}
\center
\includegraphics[width=7 in]{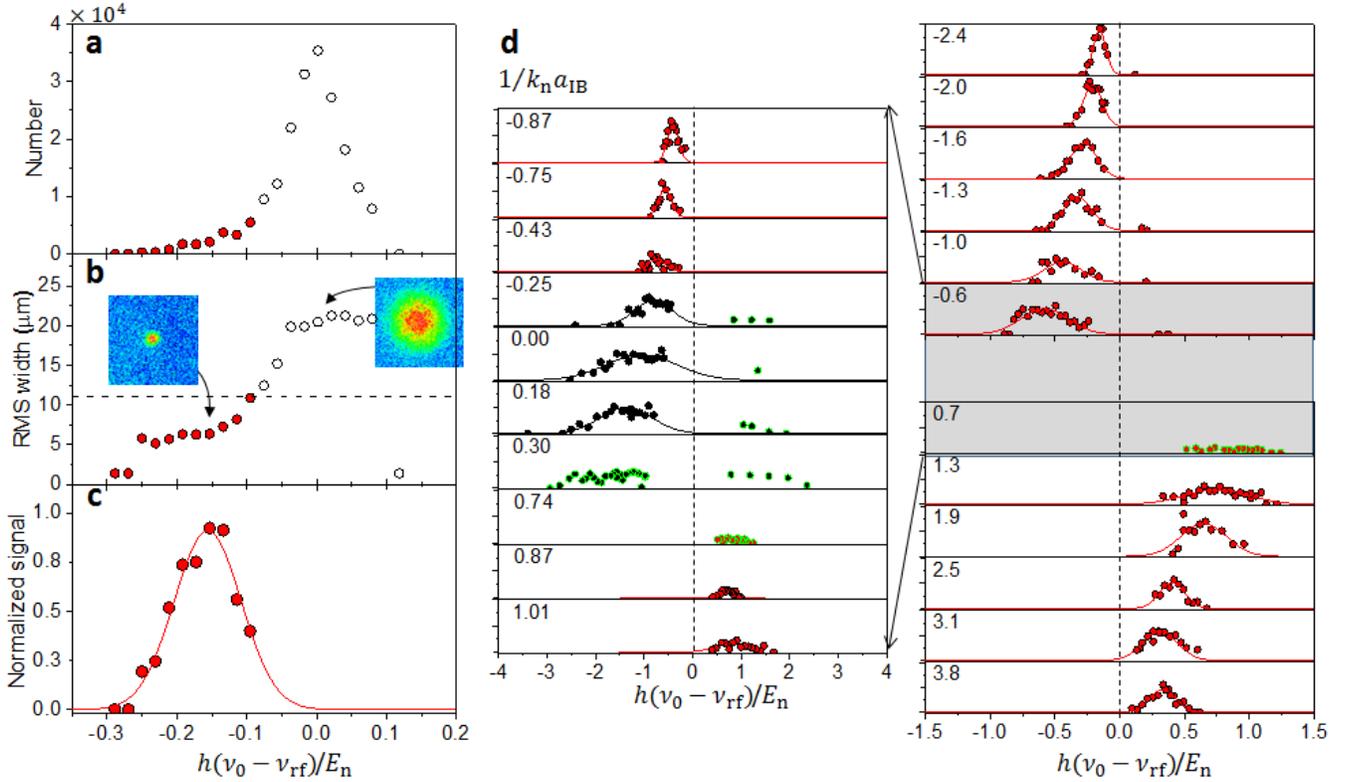}
\caption{RF spectra of $^{40}$K impurities in a $^{87}$Rb BEC.  At the relatively weak interaction of $1/k_\text{n}a_\text{IB}=-2.4$, a typical RF spectrum of impurities is shown in (\textbf{a}-\textbf{c}). \textbf{a}, The total number of atoms spin-flipped into $|\downarrow\rangle$ impurity atoms with variable interactions. \textbf{b}, the RMS width of the imaged impurity cloud. The filled points correspond to data where the width is less than $11$ $\mu$m (dashed line). The insets show absorption images at two values of $\nu_{\text{rf}}$. \textbf{c}, The normalized central density of the impurity atoms, obtained using an inverse Abel transform. The red line shows a fit to a gaussian function. \textbf{d}, RF spectra, as shown in (c), for different interaction strengths. Red and black dots show data taken with long and short RF pulses [see text], respectively. Data shown in green circles were excluded from the gaussian fits.}
\label{fig:lineshape}
\end{figure*}

In our experiment, we typically have $2\times 10^5$ Bose-condensed Rb atoms in a harmonic trap whose radial and axial frequencies are $f_{\text{Rb},\rho}=39$ Hz and $f_{\text{Rb},z}=183$ Hz, respectively. The Thomas-Fermi radii of the BEC are typically $15$ $\mu$m radially and $3.2$ $\mu$m axially. The peak density of the condensate is $n_\text{BEC}=1.8\times 10^{14}$ cm$^{-3}$. The same optical trap typically contains $2.5\times10^{4}$ K atoms, where the trap frequencies are $f_{\text{K},\rho}=50$ Hz and $f_{\text{K},z} = 281$ Hz. Our experiment usually operates at a temperature of $180$ nK, such that the fermionic impurity atoms have $T/T_F$$\approx$$0.4$, where $T_F$ is the Fermi temperature; we expect that their quantum statistics are not important. The peak potassium density is $n_\text{K}=2\times 10^{12}$ cm$^{-3}$.
The BEC is weakly interacting with $a_\text{BB}=100$ $a_0$ \cite{Kempen2002}, where $a_0$ is the Bohr radius. Following refs.~\cite{Rath2013,Li2014}, we define  momentum and energy scales, respectively, by $k_\text{n}$=$(6\pi^2\bar{n}_{\text{BEC}})^{1/3}$ and $E_\text{n}$=$\hbar^2k_\text{n}^2/2m_\text{Rb}$, where $\bar{n}_\text{BEC}$ is the average probed density. Typically $E_\text{n}/h$ and $k_\text{n}$ are $25$ kHz and $1/(900 a_0)$, respectively.  We tune the impurity-boson interactions using a broad $s$-wave Feshbach resonance for  potassium atoms in the state $|\downarrow\rangle \equiv|f=9/2,m_f=-9/2\rangle$, and $^{87}$Rb atoms in the $|1,1\rangle$ state \cite{Inouye2004}. The impurity-boson scattering length $a_\text{IB}$ as a function of magnetic field $B$ is given by $a_\text{IB}$=$a_\text{bg}[1-\delta B/(B-B_\text{0})]$, where $a_\text{bg}$=$-187\,a_\text{0}$, $B_\text{0}$=$546.62\,\mathrm{G}$, and $\delta B$=$-3.04\,\mathrm{G}$ \cite{Klempt2008}.

We measure the energy spectrum by performing RF spectroscopy. To minimize the loss due to three-body inelastic collisions, we initially prepare the impurity atoms in a state, $|\uparrow\rangle\equiv|9/2,-7/2\rangle$, that is weakly interacting with the bosons. We then apply an RF pulse at a frequency $\nu_{\text{\text{rf}}}$, detuned from the bare atomic transition $\nu_0$, to drive impurity atoms into the strongly interacting $|\downarrow\rangle$ state (see Fig. \ref{fig:Bosepolaron}\textbf{c}). Immediately after the RF pulse, we selectively image the impurity atoms in the $|\downarrow\rangle$ state. The RF photon has frequency $\nu_\text{rf}$ over a range of 79.9 to 80.5 MHz.
The RF pulse  has a gaussian envelope with the RF power proportional to $\exp(-t^2/2\Delta t^2)$. The gaussian envelope is truncated at $\pm 4\Delta t$ and the full pulse duration is $8\Delta t$.   The RF lineshape is approximately proportional to $\exp(-\nu_\text{rf}^2/2\delta\nu^2)$, where the frequency width is $\delta\nu = 1/(4\pi\Delta t)$. In the experiment, we used long $\pi$-pulses of width $\Delta t = 65$ $\mu$s ($\delta\nu = 1.2~\text{kHz}= 0.05 E_n/h$) for most of the spectroscopy data. In order to improve signal for the spectra around the unitarity regime, we used short $1.5\pi$-pulses of width $\Delta t = 12.5$ $\mu$s ($\delta\nu = 5.7~\text{kHz}= 0.23 E_n/h$).

In Fig. \ref{fig:lineshape}(a-c), we show typical RF spectroscopy data at a relatively weak interaction strength. The total  number of spin-flipped atoms (Fig. \ref{fig:lineshape}a) as a function of $\nu_\text{rf}$ is dominated by signal from $^{40}$K atoms that do not spatially overlap with the BEC (see Fig. \ref{fig:Bosepolaron}d). To distinguish signal that comes from $^{40}$K atoms that overlap with the BEC, we apply a cut based on the root-mean-square (RMS) width obtained by fitting the image to a 2D gaussian distribution (Fig. \ref{fig:lineshape}b). For those imaged clouds whose size is less than or equal to $11$ $\mu$m (red points), we apply an inverse Abel transform to extract the central density \cite{SM}. The normalized central density, averaged over a region corresponding to a radius of $R_\text{avg}=5$ $\mu$m in the transverse direction, is shown in Fig. \ref{fig:lineshape}\textbf{c}. The measured density is normalized by a calculated initial central density of the impurity atoms \cite{SM}. We use a gaussian fit (red line in Fig. \ref{fig:lineshape}c) to this spectrum to obtain the energy shift, $\Delta$, defined as the fit peak (corrected for initial interactions), and the spectral width. Figure \ref{fig:lineshape}d shows RF spectra similar to Fig. \ref{fig:lineshape}c but at different interaction strengths. The widths of almost all of the spectra remain sufficiently small that a clear peak is observed, which is consistent with a quasiparticle description of the excitation.

\begin{figure}
\includegraphics[width=3.4in]{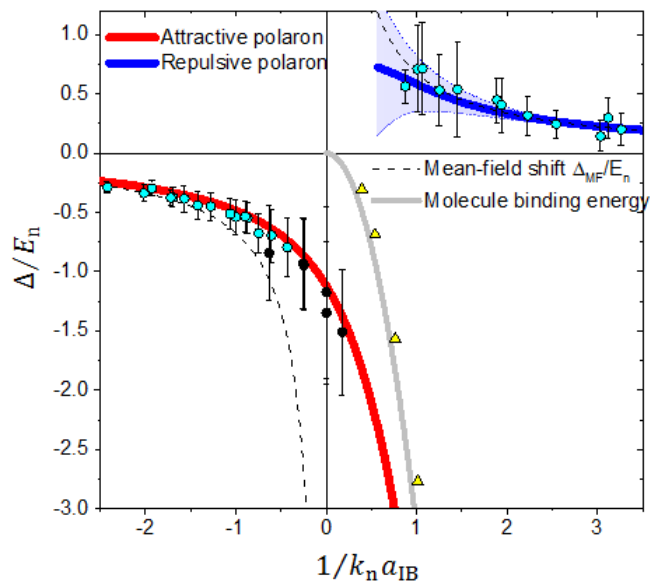}
\caption{Energy shift, $\Delta$, and spectral width of Bose polarons. Cyan and black dots show data taken with long and short RF pulses, respectively. Error bars indicate the measured RMS spectral width. The mean-field prediction is shown with dashed lines. Recent Bose polaron energy predictions \cite{Rath2013,Li2014} for the attractive and repulsive branches are shown with red and blue lines, respectively.  The gray line shows the universal two-body prediction for the energy of KRb Feshbach molecules. Triangles show separate measurements of this energy using RF association of molecules performed in a very low-density Rb gas. This two-body result is not valid for the high-density regime and is shown only for reference. The blue shaded area indicates the predicted spectral width of the repulsive branch.}
\label{fig:energy}
\end{figure}

Figure \ref{fig:energy} shows the fit energy shifts and spectral widths. To extract the polaron energy from the RF spectra, we need to account for the initial mean-field shift. The energy shift $\Delta$ is given by
\begin{equation*}\label{eq:Delta}
\Delta=h(\nu_0-\nu_{\text{p}})+E_{\text{bg}},
\end{equation*}
where $\nu_\text{p}$ is the fit peak; the background shift is given by $E_{\text{bg}}=g_{\text{bg}}\bar{n}_\text{BEC}$, where $g_{\text{bg}}=2\pi a_{\text{bg}}\hbar^2/\mu_\text{KRb}$.  For comparison, we show the predicted energy of the Bose polaron at zero temperature and momentum for the  attractive and repulsive branches as calculated in the T-matrix approach of ref. \cite{Rath2013}. Similar results have been obtained in ref. \cite{Li2014}. For weak interactions away from Feshbach resonance (large $|1/k_\text{n}a_{\text{IB}}|$), the theoretical energies are well-described by a mean-field shift given by $\Delta_\text{MF}/E_\text{n}=(2m_\text{Rb}/3\pi\mu_\text{KRb})(k_\text{n}a_\text{IB})$ (dashed lines in Fig. \ref{fig:energy}), where $m_\text{Rb}$ is the Rb mass and $\mu_\text{KRb}$ is the reduced mass between potassium and rubidium atoms. On the positive-$a_\text{IB}$ side of the resonance, the attractive branch is predicted to asymptotically approach the energy of the Feshbach molecule.  Also shown is the universal density-independent prediction for the molecule energy, $E_{\text{mol}}/E_\text{n}=-(2m_\text{Rb}/\mu_\text{KRb})(1/k_\text{n} a_\text{IB})^2$, and measurements of the binding energy of weakly bound molecules. These latter data are separately measured by  RF association of K atoms immersed in a very low-density non-condensed Rb gas, and are similar to measurements reported in refs. \cite{Klempt2008,Zirbel2008a}.

We find that the measured energy shifts match well with the predictions for the Bose polaron across the entire range of impurity-boson interactions.  In particular, our data are consistent with an attractive polaron (red line) that exists across unitarity ($1/k_\text{n}a_\text{IB}=0$) and has an energy that asymptotically approaches that of the low-density molecular state (gray line). When trying to extend the range of the data in the strongly interacting regime, we observe signal in the measured RF spectrum, but no clear peak (see, for example, the green points in Fig. \ref{fig:lineshape}d).

\begin{figure}
\includegraphics[width=3.4in]{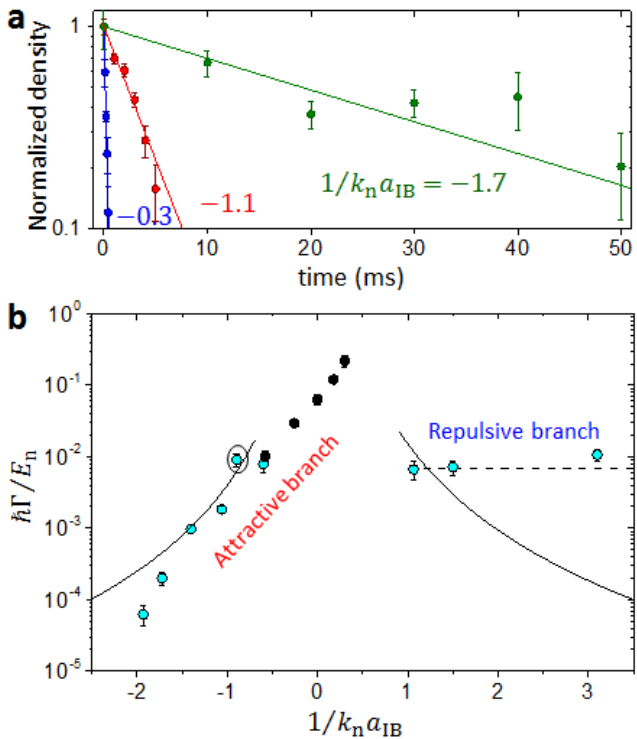}
\caption{Decay rate of the Bose polaron. \textbf{a}, Example decay rate measurements. Each point is the average of five shots with the error bar being the standard error of the mean. Lines are fits to an exponential with a decay rate, $\Gamma$. \textbf{b}, Decay rates at different interaction strengths. Cyan and black dots denote the same RF pulses as in Fig. \ref{fig:energy}. The solid lines are the predicted three-body recombination loss rate.  The dashed line show the estimated rate for atoms to leave the probed central region of the BEC: $1/(R_\text{avg}/\sqrt{k_BT/m_\text{K}})$, where $R_\text{avg}$=5 $\mu$m. The circled dot is near a narrow $d$-wave Feshbach resonance \cite{Bloom2013}, which could cause an increased decay rate.}
\label{fig:lifetime}
\end{figure}

The theory, which does not include three-body interactions, predicts a finite lifetime for the repulsive branch. This results in spectral broadening as indicated by the blue shaded region in Fig. \ref{fig:energy}. In contrast, the attractive polaron is modeled as the ground state of the system, and as such, should exhibit no lifetime broadening.  More generally, we do expect three-body loss to more deeply bound states to occur at all interaction strengths, including on the attractive branch. To measure this decay, we spin-flip the impurities into the polaron state and hold for a variable amount of time before imaging.  We choose a value of $\nu_\text{rf}$ close to the measured polaron peak and measure the central density of $|\downarrow\rangle$ atoms as a function of the hold time. Fig. \ref{fig:lifetime}a shows example lifetime measurements for the attractive polaron at three different interaction strengths. We obtain a decay rate $\Gamma$ by fitting to an exponential and the results are presented in Fig. \ref{fig:lifetime}b.  

For the attractive branch (data for $1/k_\text{n}a_\text{IB}<1$), the decay rate increases dramatically as the interaction strength is increased and continues to increase across $1/k_\text{n}a_\text{IB}=0$. For comparison, the solid line in Fig. \ref{fig:lifetime}b shows the expected three-body recombination decay rate, which scales as $(k_\text{n}a_\text{IB})^4$ \cite{D'Incao2005} and has a coefficient from previous measurements \cite{Bloom2013}. Because this scaling is not valid for the strongly interacting regime, we show the curves only for $|1/k_\text{n}a_\text{IB}|>1$. The measured upper limit for the polaron decay rate has a similar magnitude and trend as the expected loss rate for three-body recombination, which suggests that three-body inelastic processes involving an impurity atom and two bosons play an important role in the decay rate measurements. For the repulsive branch, we measure a decay rate that is independent of the interaction strength. We attribute this to dynamics that result from repulsive interactions, which expel the impurities/polarons from the center of the BEC. The dashed line shows a rough estimate of the time that it takes a K atom to leave the central region that we probe.

The measured loss rate, along with the RF pulse linewidth, contribute to the total width of the RF spectra indicated by the error bars in Fig.~\ref{fig:energy}. However, our measured widths are generally larger than the quadrature sum of these broadening effects. On the attractive (repulsive) branch, the spectra are sometimes broader by as much as a factor of 3 (8). Measured widths increase for stronger interaction strengths (smaller $|1/k_\text{n}a_\text{IB}|$) and are larger for the repulsive branch than for the attractive branch. While the spectral width contains information about the polaron lifetime, there are several experimental factors that can also contribute to broadening. We estimate that density inhomogeneity introduces broadening by up to 15 percent of the shift at the center of the polaron peak, $\Delta$. Beyond this, the RF spectra could additionally be broadened by intrinsic many-body effects and the thermal momentum distribution of impurities. It must be acknowledged, however, that if this unaccounted-for width gives rise to asymmetric lineshapes, there will be systematic errors in our energy determination.  The excess broadening, in any case, gives a lower limit on the polaron lifetime, which is still longer than the quasiparticle's inverse energy. We conclude that the Bose polaron is a well-defined quasiparticle.

This first measurement of the Bose polaron in a three-dimensional trapped atom gas probed the energies and lifetimes for both the attractive and repulsive branches. A future direction would be to probe other basic quasiparticle properties, such as the quasiparticle's effective mass and residue. In addition, it would be interesting to explore the dependence of the excitation spectrum on temperature, both theoretically and experimentally. In our case, by probing impurities in the central region of the BEC, the effective temperature of the bosons should be very close to zero. Finally, ref. \cite{Levinsen2015} theoretically considers the effect of Efimov trimers on the Bose polaron. However, for our system, the location of the Efimov resonance is predicted to be $a_-<-30,000a_0$ \cite{Wang2015}, such that Efimov effects on the Bose polaron would only occur in an extremely narrow regime of small $|1/k_\text{n}a_\text{IB}|$ that we do not resolve. Experimental investigation of Efimov effects on the Bose polaron would thus require a different atomic system with a more favorable three-body interaction parameter.

As this paper was being readied for submission, we learned of interesting parallel experimental work \cite{Jorgensen2016}.

\begin{acknowledgements}
This work is supported by NSF under Grant No. 1125844, and by NASA under Grant No. NNN12AA01C.
\end{acknowledgements}


\section{Supplemental Material}

\textbf{Inverse Abel transform.} The inverse Abel transform can be used to calculate the three-dimensional density distribution from a two-dimensional absorption image when the system has ellipsoidal symmetry. However, differential gravitational sag ($\sim 4$ $\mu$m) between the K cloud and the Rb BEC breaks this symmetry. We have performed a numerical simulation to show that the inverse Abel transform nevertheless works well for extracting the RF spectrum of Bose polarons at the BEC center as long as the signal comes predominantly from impurity atoms inside the BEC. These atoms approximately share the ellipsoidal symmetry of the condensate due to the initial attraction between $|\uparrow\rangle$ impurities and the BEC. (Recall that $a_\text{bg}=-187a_0$.)

In our simulation, we use a Thomas-Fermi profile for the BEC density, $n_\text{Rb}(\rho,z)$, and calculate the initial K atom density distribution using
\begin{eqnarray}\label{eq:density}
n_{\uparrow}(\rho,z)&=&n_\text{K0} \exp \left[-\rho^2/2\sigma_\rho^2-(z+z_\text{sag})^2/2\sigma_z^2\right.\nonumber\\
&&\left.-g_{bg}n_\text{Rb}(\rho,z)/k_\text{B}T\right],
\end{eqnarray}
where  $\rho=\sqrt{x^2+y^2}$, $\sigma_{\rho,z}=\sqrt{k_\text{B}T/m_\text{K}}/(2\pi f_{\text{K},\rho,z})$, and $k_\text{B}$ is the Boltzmann constant. Here, we assume thermal equilibrium and ignore the quantum statistics of the fermionic $^{40}$K atoms. The RF spectroscopy is modeled based on mean-field theory to give an impurity cloud density of
$n_{\downarrow}(\rho,z,\nu_\text{rf})= n_{\uparrow}(\rho,z)\exp\left[-\left(\nu_\text{rf}-g_\text{bg} n_\text{Rb}(\rho,z)\right)^2/2\delta\nu^2\right]$,
where we set the bare atom transition to $\nu_0=0$. We can then compare the central density of the impurity cloud to that returned by calculating a simulated absorption image (integration along $z$). In this comparison, we include the effect of our measured imaging resolution of 3.4 $\mu$m by convolving the density distribution and absorption image with a gaussian function.

From the simulation, we find that the inverse Abel transform returns the correct central density of the $|\downarrow\rangle$ atoms for values of $\nu_\text{rf}$ where the imaged atom cloud has an RMS width less than or equal to $11$ $\mu$m. The simulated RF spectrum, using the inverse Abel transform, also allows us to estimate the probed BEC density. We find that the energy shift of the peak of the RF spectrum corresponds to $\bar{n}_\text{BEC}=0.85n_\text{BEC}$. 

The measured central densities in Fig. 2(c-d) (obtained using the inverse Abel transform) are normalized by the calculated central density of the initial $|\uparrow\rangle$ impurity atoms based on Eq. \ref{eq:density} and including our imaging resolution. The measured $^{40}$K atom number, number of atoms in the BEC, and $T$ extracted from fitting the tail of the initial $^{40}$K density distribution are inputs to this calculated central density.

\end{document}